\documentclass[conference]{IEEEtran}

\usepackage{cite}
\usepackage{amsmath,amssymb,amsfonts}
\usepackage{algorithmic}
\usepackage{graphicx}
\usepackage{textcomp}
\usepackage{xcolor}
\usepackage{url}

\def\BibTeX{{\rm B\kern-.05em{\sc i\kern-.025em b}\kern-.08em
    T\kern-.1667em\lower.7ex\hbox{E}\kern-.125emX}}

\usepackage{listings}
\usepackage{xspace}
\usepackage{subfig}
\usepackage{booktabs}

\newcommand{\pipit}{Pipit\xspace}

\definecolor{MyRed}{HTML}{CF0000}
\definecolor{MyGreen}{HTML}{009E73}
\newcommand{\grcheck}{{\color{MyGreen}\checkmark}}
\newcommand{\rdcross}{{\color{MyRed}$\times$}}

\definecolor{dkgreen}{RGB}{0,64,0}
\definecolor{ltgray}{RGB}{245,245,245}
\definecolor{mauve}{RGB}{139,0,139}

\definecolor{codegreen}{rgb}{0,0.6,0}
\definecolor{codegray}{rgb}{0.5,0.5,0.5}
\definecolor{codepurple}{rgb}{0.58,0,0.82}
\definecolor{backcolour}{rgb}{0.94,0.97,1}

\lstdefinestyle{mystyle}{
    language=Python,
    backgroundcolor=\color{backcolour},   
    commentstyle=\color{codegreen},
    keywordstyle=\color{magenta},
    numberstyle=\tiny\color{codegray},
    stringstyle=\color{codepurple},
    basicstyle=\ttfamily\scriptsize,
    breakatwhitespace=false,         
    breaklines=true,                 
    captionpos=b,                    
    keepspaces=true,                 
    numbers=left,                    
    numbersep=5pt,                  
    showspaces=false,                
    showstringspaces=false,
    showtabs=false,                  
    tabsize=2,
    stepnumber=1,
    rulecolor=\color{black}
}

\begin{document}

\title{\pipit: Scripting the analysis of parallel execution traces}

\author{\IEEEauthorblockN{Abhinav Bhatele, Rakrish Dhakal, Alexander Movsesyan, Aditya K.~Ranjan, Onur Cankur}
\IEEEauthorblockA{~\\
    \textit{Department of Computer Science}\\
    \textit{University of Maryland}\\
    E-mail: bhatele@cs.umd.edu, \{rakrish, amovsesy, aranjan2, ocankur\}@umd.edu, 
}
}

\maketitle

\begin{abstract}
Performance analysis is a critical step in the oft-repeated, iterative process
of performance tuning of parallel programs. Per-process, per-thread traces
(detailed logs of events with timestamps) enable in-depth analysis of parallel
program execution to identify different kinds of performance issues. Often
times, trace collection tools provide a graphical tool to analyze the trace
output. However, these GUI-based tools only support specific file formats, are
challenging to scale to large trace sizes, limit data exploration to the
implemented graphical views, and do not support automated comparisons of two or
more datasets. In this paper, we present a programmatic approach to analyzing
parallel execution traces by leveraging pandas, a powerful Python-based data
analysis library. We have developed a Python library, \pipit, on top of pandas
that can read traces in different file formats (OTF2, HPCToolkit, Projections,
Nsight Systems, etc.) and provides a uniform data structure in the form of a
pandas DataFrame. \pipit provides operations to aggregate, filter, and
transform the events in a trace to present the data in different ways. We also
provide several functions to quickly and easily identify performance issues in
parallel executions. More importantly, the API is easily extensible to support
custom analyses by different end users.

\end{abstract}

\begin{IEEEkeywords}
performance analysis, traces, scripting
\end{IEEEkeywords}


\section{Motivation}
\label{sec:motiv}
Software development in parallel and high performance computing (HPC) often
involves an iterative process of writing code, analyzing performance to
identify issues, making code changes to tune performance, and then doing more
analysis and tuning. Hence, it is important to optimize the performance
analysis workflow to save developer time and effort as much as possible.
Detailed performance analysis, and specifically tasks such as critical path
detection, computation-communication overlap analysis, and root cause analysis,
often require the collection and analysis of parallel execution traces.
Execution traces are detailed logs of individual events (compute, communication
routines, I/O etc.) with timestamps. Several performance tools such as
Score-P~\cite{scorep}, HPCToolkit~\cite{hpctoolkit}, Nsight
Systems~\cite{nsys}, and Projections~\cite{namdPerfFGCS} can collect
per-process, per-thread and even per-GPU traces of parallel programs.

When production scientific applications are launched with a large number of
processes and/or threads even for short periods of time, the traces grow in
size and complexity quickly due to the large number of events logged per
process/thread. This often makes the task of trace analysis and visualization
unwieldy and challenging. Several trace collection tools also provide a
corresponding graphical tool for analyzing trace data -- some popular examples
are Vampir~\cite{vampir}, hpcviewer~\cite{mellorcrummey+:jsc02,
adhianto2010effectively}, Jumpshot~\cite{jumpshot} and Nsight
Systems~\cite{nsys}. However, most GUI-based tools have some limitations.
First, each tool only supports specific file formats, and as a result, end
users have to familiarize themselves with the interfaces of multiple tools to
effectively analyze traces in different formats. Second, when using GUI-based
tools, end users are constrained in their exploration of the data by the views
provided by each tool. Most visualization tools are not extensible easily to
support new kinds of analyses. Moreover, in most graphical tools, repeating the
same analysis twice on the same or different datasets is a manual process, with
limited support for saving/automating analysis. Third, since parallel traces
can be large in two dimensions (time and number of processes/threads),
visualizations often have issues with scalability beyond small datasets.
Finally, while some graphical tools allow loading trace data from two different
executions, the exploration is user-driven and manual, and there is often
limited or non-existent support for automated comparisons of two or more
datasets.

There are several challenges/requirements to developing a tool for trace
analysis that solves all of the issues mentioned above. It should be able to
handle different file formats for execution traces. It should allow for canned
and custom exploration of the data by the end user. Further, it should enable
automating performance analysis for oft-repeated tasks. And, it should allow
comparisons of traces from multiple executions, hopefully in a semi-automated
or scriptable manner.

\setlength\tabcolsep{4pt}
\begin{table*}[t]
\caption{Capabilities in different trace visualization and analysis tools, including this work.}
\label{tab:survey}
\begin{tabular}{lccccccccccccc} \toprule
            & Interface& Events    & Metrics   & Call  & Flat    & Time    & Outlier  & Comm.  & Msg Size  & Pattern & Manual    & Guided    &Extensible\\
            & type     & over time & over time & Stack & Profile & Profile & Analysis & Matrix & Histogram & Detect. & Mult.~Run & Mult.~Run & API      \\ \midrule
Vampir      & GUI      & \grcheck & \grcheck & \grcheck & \grcheck & \grcheck & \rdcross & \grcheck & \grcheck & \rdcross & \grcheck & \rdcross & \rdcross \\
hpcviewer   & GUI      & \grcheck & \rdcross & \grcheck & \grcheck & \grcheck & \rdcross & \rdcross & \rdcross & \rdcross & \rdcross & \rdcross & \rdcross \\
Projections & GUI      & \grcheck & \grcheck & \rdcross & \rdcross & \grcheck & \grcheck & \grcheck & \grcheck & \rdcross & \grcheck & \rdcross & \rdcross \\
Nsight Sys. & GUI      & \grcheck & \grcheck & \grcheck & \grcheck & \rdcross & \grcheck & \rdcross & \rdcross & \rdcross & \grcheck & \rdcross & \rdcross \\
Perfetto    & GUI, SQL & \grcheck & \grcheck & \grcheck & \grcheck & \rdcross & \rdcross & \rdcross & \rdcross & \rdcross & \rdcross & \rdcross & \grcheck \\
Scalasca    & GUI, CLI & \rdcross & \grcheck & \grcheck & \rdcross & \rdcross & \grcheck & \rdcross & \grcheck & \grcheck & \rdcross & \rdcross & \rdcross \\ \midrule
This work   & Python   & \grcheck & \grcheck & \grcheck & \grcheck & \grcheck & \grcheck & \grcheck & \grcheck & \grcheck & \grcheck & \grcheck & \grcheck \\ \bottomrule
\end{tabular}
\end{table*}

In this paper, we fill the above mentioned gaps in performance analysis of
parallel execution traces by developing a Python-based API to analyze them.
This API provides full access to the trace data to the end user so that they
can explore the data programmatically instead of relying on a graphical
interface. Since trace data consists of a time series of events (with
categorical and numerical attributes for each event), we leverage
pandas~\cite{mckinney:pandas}, a powerful Python-based library for analyzing
tabular and time series data. We have implemented our API in a performance
analysis library called \pipit (name anonymized for double-blind review) that
reads traces from different file formats (OTF2~\cite{eschweiler2012open},
HPCToolkit~\cite{hpctoolkit}, Projections~\cite{namdPerfFGCS}, Nsight
Systems~\cite{nsys}, etc.) and provides uniform data structures built atop the
pandas DataFrame. \pipit exposes an API with operations to aggregate, filter,
and transform the events in a trace dataset to explore, manipulate, and
visualize the data in different ways. \pipit implements both low-level
functions to manipulate trace data directly, as well as high-level functions
for easy scripting of common analyses by end users. Users can also optionally
access the underlying DataFrame to perform custom data wrangling operations
that are specific to their traces, which may not be already implemented in
\pipit.

There are several common data exploration/manipulation tasks that end users
perform when analyzing parallel traces. Some examples are -- analyzing a heat
map or matrix of communication between MPI processes, detecting load imbalance
across threads or processes, detecting a critical path in the execution,
identifying the degree of overlap of communication with computation, etc. We
have designed and implemented many of these operations in the \pipit API to
reduce end user effort for such common performance analysis tasks. We also
present several performance analysis case studies that demonstrate the utility
and capabilities of \pipit.

The paper makes the following important contributions:
\begin{itemize}
\item A unified Pythonic interface to read traces generated in different file
formats by different trace collection tools.
\item An open-source library, \pipit that provides a programmatic interface to
perform common performance analysis tasks with ease using implemented
functions.
\item Basic visualization support with tens of graphical views to complement
the \pipit API in assisting with performance analysis.
\item Demonstration of the utility of \pipit in identifying performance issues
in several HPC applications.
\end{itemize}

\section{Background and Related Work}
\label{sec:bg}
We first describe what an execution trace looks like, and mention some trace
collection tools, and then briefly discuss trace visualization and analysis
tools.

\subsection{Execution Traces and Trace Collection Tools}

An execution trace for an individual process or thread is a time-series of
events that occur within a program's execution on that process/thread
(representing function calls, loops, and other code blocks). Each event
contains a timestamp, the type of event, and other optional metrics such as
values of hardware counters. A trace of a parallel program contains this
information for multiple processes and/or threads executing within a program
execution.

There are many popular trace collection tools such as Score-P~\cite{scorep},
HPCToolkit~\cite{hpctoolkit}, TAU~\cite{shende:tau2006}, Nsight
Systems~\cite{nsys}, Projections~\cite{namdPerfFGCS}, and PyTorch Profiler
~\cite{pytorch-profiler}. Score-P, HPCToolkit, and TAU are general purpose
tracing tools that can collect trace data for any C/C++/Fortran program. Nsight
Systems is used for GPU-enabled programs running on NVIDIA GPUs, and
Projections is specific to Charm++~\cite{CharmppOOPSLA93} programs. Finally,
PyTorch Profiler collects trace data during training and inference for
PyTorch-based deep learning programs.

\subsection{Trace Visualization Tools}

The tracing tools mentioned above are developed by different groups, and use
different file formats and serialization techniques to store the parallel trace
data. Most of them also provide complementary visualization tools to visualize
the traces. As we began development of \pipit, we conducted a survey of
existing trace visualization and analysis tools, and their strengths and
weaknesses. Table~\ref{tab:survey} provides a summary of our study. For each
tool in consideration, we evaluated if a certain graphical view that represents
a certain kind of analysis of trace data was available or not. Below, we
provide a high-level overview of each of the tools we considered.

Vampir~\cite{vampir}, hpcviewer~\cite{mellorcrummey+:jsc02,
adhianto2010effectively}, Projections~\cite{namdPerfFGCS}, and Nsight
Systems~\cite{nsys} are GUI-based tools offering various views for trace
analysis.  All of these GUI-based tools offer a timeline view for visualizing
trace events over time. Most also provide views for metric values over time,
call stack views, and statistical views, as well as different levels of support
for filtering by attributes and time range. None of these tools offer a {\it
programmable} interface for dealing with raw execution trace data from parallel
runs. Users must point and click to analyze the data, which can be time
consuming and inflexible for large datasets or custom analyses.

As seen in Table~\ref{tab:survey}, some offer a flat profile view, which shows
the time spent in each function, as well as a time profile view, which shows
how the flat profile changes over time. Projections additionally offers
object-specific information for Charm++ program tasks. Only Vampir and
Projections support communication matrix views, whereas none of these tools
support any type of pattern detection or guided comparative analysis of
multiple runs.  Finally, there are other visualization-based tools which we did
not analyze in detail, such as Jumpshot~\cite{jumpshot}, AMD's
OmniTrace~\cite{omnitrace} and Paraver~\cite{paraver}.  Again, these are GUI
tools and they do not provide the flexibility to easily script new analyses or
to easily query, filter, or aggregate trace data in an indexed DataFrame as
\pipit does.  Typically, the available analyses are manually selected through
drop down menus or some other user-interface, and there is limited flexibility
for customization.

\subsection{Other Related Analysis Tools}

Scalasca~\cite{geimer2010scalasca}, unlike interactive, GUI-based tools,
analyzes OTF2 traces and generates a report. This report contains occurrences
of patterns that Scalasca searches for in the trace, which are linked to
performance bottlenecks.  The generated report can be inspected using the
Cube~\cite{saviankou2015cube} program, which offers both a command-line
interface and a GUI. HTA~\cite{hta} is a Python-based tool that is designed
specifically for analyzing PyTorch traces.  It provides features that are
specific to PyTorch applications, such as kernel breakdown and showing GPU
performance counters.  Perfetto~\cite{perfetto}, although not developed for HPC
applications, provides a robust timeline viewer and a SQL engine for querying
trace data.  However, it requires conversion of trace data to a supported
format, as it doesn't support common HPC trace formats.

An alternative to detailed tracing is sampling-based profiling, which generates
aggregate information. Various tools exist for capturing and analyzing these
profiles. For instance, Hatchet~\cite{bhatele:sc2019} provides a Python
interface that enables programmatic analysis of profiles, and
ParaProf~\cite{bell:paraprof2003} provides a command-line and graphical
interface for profile analysis. However, they do not support analyzing traces.

\section{The \pipit Library}
\label{sec:pipit}
We present \pipit, a Python library for programmatically analyzing parallel
execution traces. Our goals in developing this library were the
following:~(1)~Support several file formats used for collecting traces, to
provide users with a unified interface that works with outputs of many
different tracing tools;~(2)~Provide a programmatic API, which allows users to
write simple code for trace analysis, providing benefits like flexibility of
exploration, extensibility, reproducibility, and automation/saving of
workflows; and~(3)~Automate certain common performance analysis tasks for
analyzing single and multiple executions. We now describe the considerations in
designing and implementing \pipit.

\subsection{The trace as a pandas DataFrame}

As described in Section~\ref{sec:bg}, a parallel trace consists of a series of
events, with timestamps and other attributes, for each process and thread that
the program runs on. This data is inherently high-dimensional, given the
combination of (events, timestamps) $\times$ (processes, threads, GPUs)
$\times$ (performance metrics). We determined that we can treat this data as
two-dimensional by considering the combination of event, timestamp, and process
ID (rank) as one axis, and all the data collected per event, both numeric and
categorical, as the other axis. This enables us to use pandas DataFrames as the
primary data structure for organizing trace data. A pandas DataFrame is a
two-dimensional tabular data structure that allows storing both heterogeneous
and sparse data in memory~\cite{mckinney:pandas}.

In \pipit, each event is represented as a row, and each attribute as a column.
Since pandas stores the DataFrame in a column-major format, the values in each
column are stored contiguously in memory, allowing analysis operations over
rows of individual columns to be vectorized. This vectorization significantly
accelerates the manipulation, aggregation, and exploration of trace events and
metrics. Additionally, the pandas API is relatively easy to learn and use,
which lets users implement custom data wrangling operations on top of \pipit's
API. Finally, the Python software ecosystem has a large number of open-source,
third-party libraries, such as NumPy~\cite{harris2020array},
Matplotlib~\cite{Hunter:2007}, and various machine learning frameworks. This
enables the use of these libraries for further downstream tasks in conjunction
with \pipit to explore trace data. We demonstrate such use-cases by
implementing operations such as pattern detection and visualization using
open-source libraries in the \pipit API (detailed in Section~\ref{sec:api} and
Section~\ref{sec:vis}).

\begin{figure}[h]
  \centering
    \includegraphics[width=0.41\columnwidth]{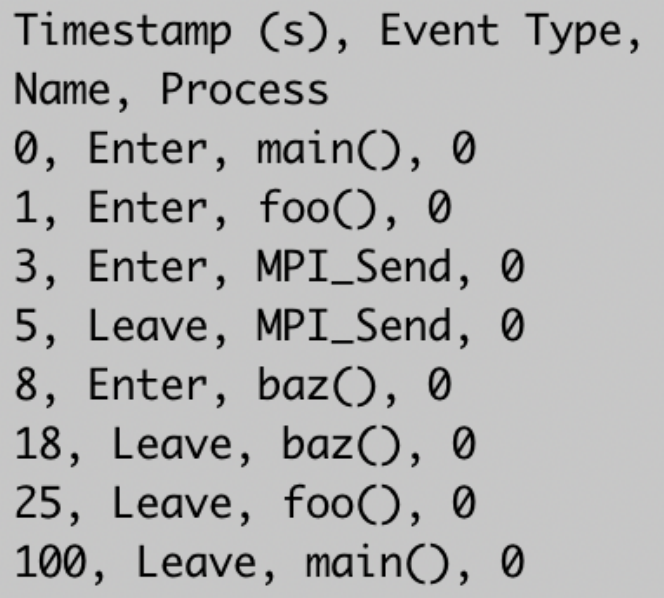}
    \hfill
    \includegraphics[width=0.56\columnwidth]{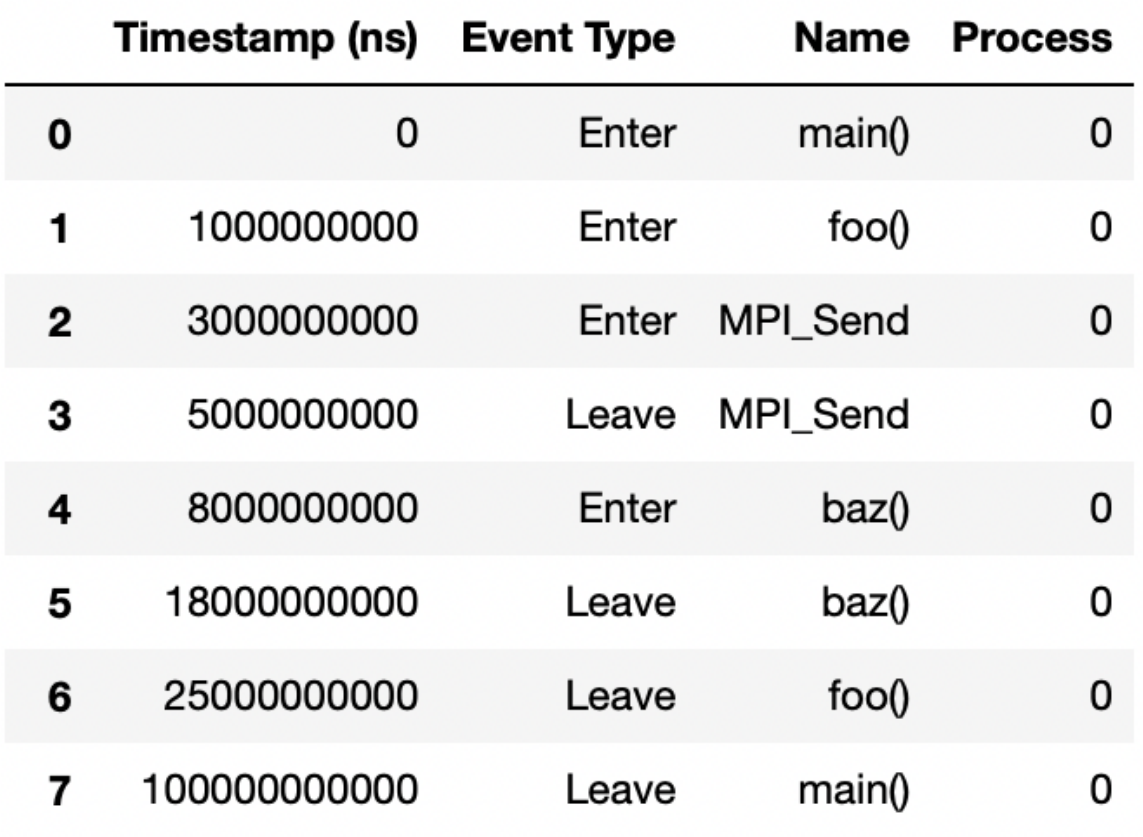}
\begin{lstlisting}
foo_bar = finch.Trace.from_csv('foo-bar.csv')
display(foo_bar.events)
\end{lstlisting}
  \caption{A sample trace file in CSV format (left), and the corresponding events DataFrame generated by \pipit after reading it (right) using the code snippet at the bottom.}
  \label{fig:toy-example}
\end{figure}

\pipit reads an execution trace into a {\tt Trace} object, which contains a
DataFrame of events. The left gray box in Figure~\ref{fig:toy-example} shows a
portion of a sample 2-process trace in CSV format.  Most tools record function
calls as a pair of events, one representing the start of a function call
(Enter) and the other when the call returns (Leave).  Each row has a timestamp
for when that event was recorded on a given process.  The image to the right
shows the corresponding DataFrame created by \pipit.  The Python code used to
generate a Trace object, {\tt foo\_bar} from an input CSV file is shown at the
bottom.

\subsection{Reading a dataset}

One of the main considerations in developing \pipit was to support various file
formats that represent different tracing tools. In order to do this, we
implemented readers for various file formats. We currently support traces in
the OTF2, HPCToolkit, Projections, Nsight Systems, and Chrome Trace Viewer
(including PyTorch) formats. \pipit's readers are uniquely designed to parse
all types of trace data into a uniform data model, while retaining all of the
original information collected by the tracing tool. This uniform data model
enables users to write single-source code that works with traces collected by
different tools in different formats.

\subsection{Generating a call graph}

Calling contexts can be useful to identify the root causes of performance
issues.  While some tracing tools such as HPCToolkit explicitly record the call
stack for each function call, most tools do not. However, by virtue of having
timestamps for each function call, we can use the nesting of these calls to
extract caller-callee relationships from the data. We use this information to
reconstruct the calling context or call stack for each function call. Given the
call stack for each event in the trace, we have several options on how to
organize this data. If we create a prefix tree from all the call stacks, we can
generate a calling context tree (CCT) at every instant in time and for every
process or thread.  However, this data would grow extremely quickly for complex
applications running on large numbers of processes. As a result, we made a
decision to aggregate the CCT along two dimensions -- over time and across all
the processes and threads. We keep this CCT in a tree data structure that is a
union of all the individual CCTs in these two dimensions.

\section{The \pipit API}
\label{sec:api}
We now describe certain operations implemented in \pipit that enable the end
user to dissect trace data in various ways. These operations help the end user
in performance analysis by making it easier, quicker, and more automated to a
certain degree. We also provide some visualization capabilities that complement
some programmatic operations (detailed in Section~\ref{sec:vis}). However,
\pipit's primary strength lies in its ability to programmatically analyze
traces.

As mentioned in Section~\ref{sec:bg}, we examined the capabilities of existing
GUI-based tools to better understand common performance analysis tasks
performed by HPC users. We have incorporated a variety of these operations into
the \pipit API as Python functions. While the \pipit API consists of many
user-facing functions, we describe a few core operations that provide insight
into the kinds of analyses \pipit can enable. To use these operations, the user
must first read in a trace dataset into a Trace object, and then apply
operations on that object.

\subsection{Extracting caller-callee relationships}

The calling contexts of function invocations can be useful to derive various
metrics that can be used in higher level analysis methods. To identify calling
contexts for each function, we first match the {\tt enter} and {\tt leave}
events that represent a function call, and then use consecutive {\tt enter}
events to identify nested function calls, deriving parent-child relationships.
The functions below can be used to automatically build the call stack per
function and the CCT for the entire program:

\vspace{0.08in}
\noindent{\bf \_match\_caller\_callee}:
The calling context of a function call is useful in context-aware performance
analysis, such as determining which call paths are responsible for certain
performance issues. These relationships are also necessary for calculating the
exclusive and inclusive times spent in each function call. We iterate through
the events and maintain a call stack for each process and thread. Using this
call stack information, we create new columns in the DataFrame that store the
parent and children relationships between functions.

\vspace{0.08in}
\noindent{\bf \_create\_cct}:
When \pipit reads in a trace, it creates a calling context tree (CCT) by
default. The CCT is stored as a separate object in the Trace object alongside
the DataFrame. Each event in the DataFrame stores a reference to its
corresponding node in the CCT. Since we store a single unified CCT across
processes and threads, it can be used to analyze discrepancies in same call
paths across different processes and help uncover any related performance
issues.

\subsection{Analyzing summary performance}

Next, we discuss API functions that summarize/aggregate detailed trace data,
and help analyze the time spent in different parts of the code.

\vspace{0.08in}
\noindent{\bf calc\_inc\_metrics} and {\bf calc\_exc\_metrics}: Since the raw
trace data has two rows per function ({\tt enter} and {\tt leave}), we need to
first calculate the time spent in each function using these two rows.  We first
match {\tt enter} and {\tt leave} events using a call stack, and then used the
matched rows to calculate the inclusive values for each function call. Then, we
use the computed parent-child relationships to compute the exclusive values by
subtracting the children's values from the parent's, giving us the exclusive
time spent, and other exclusive metrics, for the parent call.

\vspace{0.08in}
\noindent{\bf flat\_profile}:
A flat profile is often used to get a high-level overview of the most time
consuming functions in an execution. Once we compute the inclusive and
exclusive metrics per function call, we can calculate the total time spent in
each function aggregated across the entire trace. We can use the output of this
operation to focus on a subset of functions in downstream tasks.

\vspace{0.08in}
\noindent{\bf time\_profile}:
Manually inspecting the detailed timeline of a program execution with a large
number of events and processes is a laborious task. Instead, we can first
explore the activity or utilization of all the processes over time. We call
this a time profile, which provides a succinct view of the total time spent in
different functions over time across all processes. The time period captured in
the trace is divided into equal-sized time bins, and for each time bin, this
operation computes the total amount of time spent in different functions (added
across all threads and processes). One can think of this as a flat profile over
time. 

\begin{figure}[h]
    \centering
    \includegraphics[width=\columnwidth]{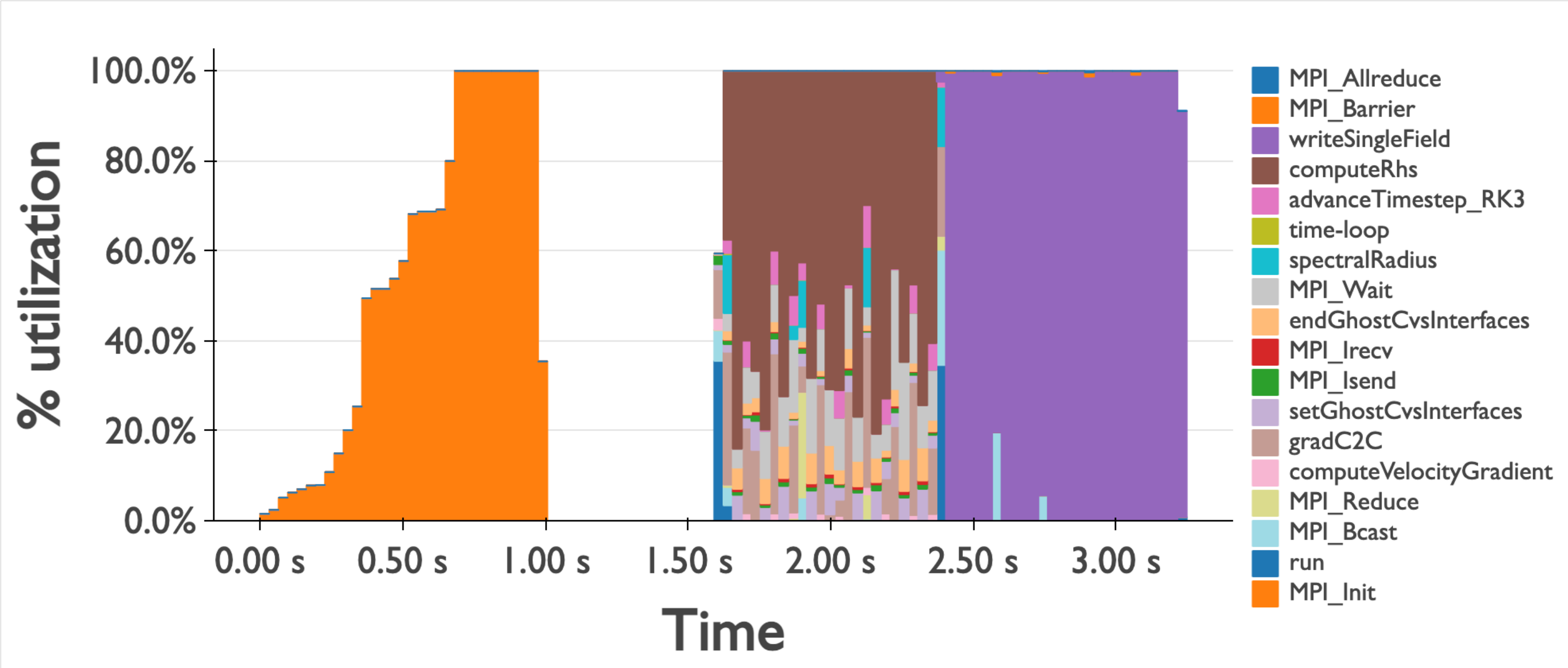}
    \caption{Time profile of a Tortuga trace with 64 processes.}
    \label{fig:time-profile-2}
\end{figure}

In Figure \ref{fig:time-profile-2}, we can see the time profile of a
computational fluid dynamics code, Tortuga, running on 64 processes. The
visualization uses a function from the plotting API described in
Section~\ref{sec:vis}. The stacked bar chart allows the user to see what
functions take up the most amount of time in a specific bin. Focusing on the
middle of the time profile, we observe that the {\tt computeRhs} function (in
brown) makes up a significant portion of the total time.

\subsection{Analyzing communication performance}

Communication is often a scalability bottleneck in message passing programs.
Hence, \pipit provides a variety of functions to script the analysis of
communication in parallel programs.  Note that only a subset of trace
collection tools record communication data, so the availability of a certain
operation depends on the data present in the trace.

\begin{figure}[h]
  \centering
  \includegraphics[width=0.49\columnwidth]{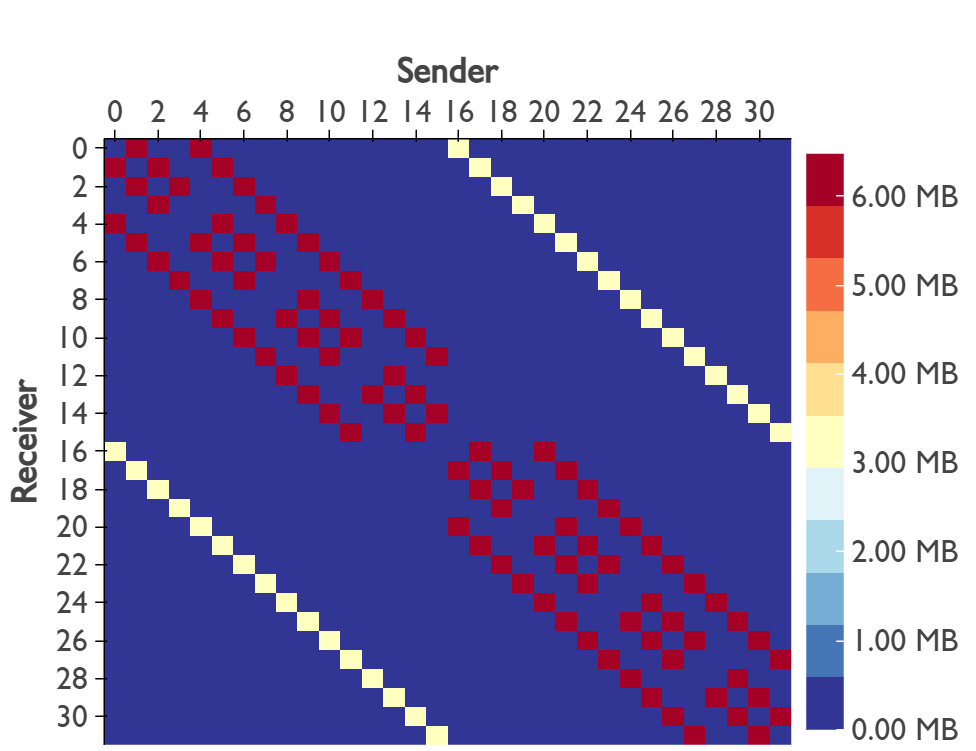}
  \includegraphics[width=0.49\columnwidth]{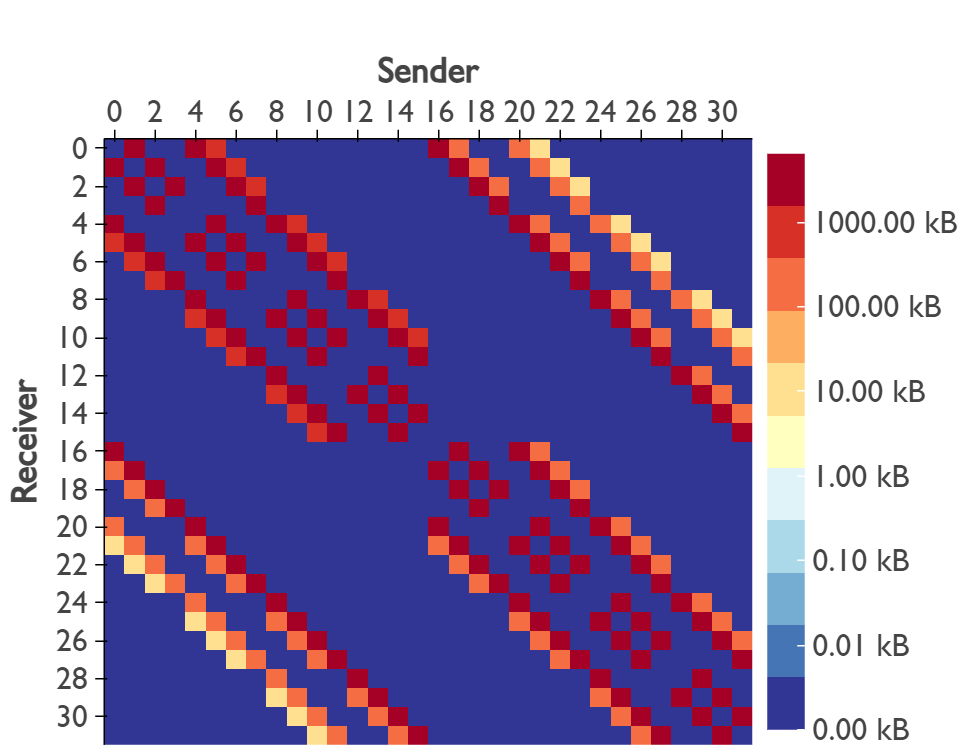}
\begin{lstlisting}
laghos_32 = finch.Trace.from_otf2('./laghos_32')

laghos_32.plot_comm_matrix(cmap='linear')
laghos_32.plot_comm_matrix(cmap='log')
\end{lstlisting}
  \caption{Communication matrix of a Laghos execution on 32 processes, with a
linear colormap (left) and logarithmic colormap (right).}
  \label{fig:laghos_32_comm_matrix}
\end{figure}

\vspace{0.08in}
\noindent{\bf comm\_matrix}:
A communication matrix represents the amount of data exchanged between pairs of
processes. The user can choose to analyze the total number of messages or total
volume of communication between each process-pair. This function outputs a
two-dimensional array, which can then be visualized using any Python libraries.
Figure~\ref{fig:laghos_32_comm_matrix} shows the communication matrix of a
Laghos execution on 32 processes, using linear and logarithmic colormaps. We
use \pipit's complementary visualization capabilities to show this matrix as a
heatmap (more details in Section~\ref{sec:vis}). We observe that the matrix is
symmetric, and the communication happens along the diagonal. This suggests a
near-neighbor communication pattern in an $n$-dimensional virtual topology.
Switching to logarithmic scale for the colormap reveals more subtle
communication patterns.

\begin{figure}[h]
    \centering
    \includegraphics[width=\columnwidth]{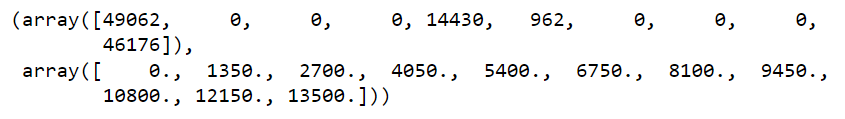}
    \caption{Message size histogram of a Laghos execution on 32 processes. We see that messages are not distributed uniformly.}
    \label{fig:laghos_32_message_histogram}
\end{figure}

\vspace{0.08in}
\noindent{\bf message\_histogram}
returns a distribution of the sizes of all messages encountered in the trace.
This can help answer questions such as -- are there a large number of small
messages, or low numbers of large messages? Figure
\ref{fig:laghos_32_message_histogram} shows this for a 32-process Laghos trace.
We see that the messages in this trace are clustered into three size ranges:
small messages (0-1,350 bytes), medium messages (5,400-6,750 bytes), and large
messages (12,150-13,500 bytes).

\vspace{0.08in}
\noindent{\bf comm\_by\_process}
returns the total message volume sent and received by each process. This lets
us quickly observe high-level communication discrepancies or imbalances across
processes in a multi-process execution.  Similar to {\tt comm\_matrix}, we can
choose to look at the number of messages or communication volume.

\vspace{0.08in}
\noindent{\bf comm\_over\_time}:
The previous operations generate statistics for communication that are
aggregated over time. However, since the trace has timestamps, we can also
analyze the messaging behavior of a program over time. This operation
calculates both the number of messages, and the total message volume, sent over
different time bins. This can gives us an idea of the spread of communication
over time and if it is bursty or uniform.

\vspace{0.08in}
\noindent{\bf comm\_comp\_breakdown}: Many communication libraries such as MPI
and NCCL now support asynchronous point-to-point and collective operations.
Often times, a programmer wants to know how much of the communication are they
able to overlap with useful computation.  This operation returns the aggregated
amount of time spent in one of four things: non-overlapped computation,
computation overlapped with communication, non-overlapped communication, and
other other functions. While one can visually see such overlap in timeline
views, most GUI-based tools do not enable quantify the degree of overlap
between computation and communication aggregated across the entire trace. We
believe that the scripting functionality provided by this function can be
extremely useful as we will demonstrate in a case study in
Section~\ref{sec:multirun}.

\subsection{Identifying performance issues}

Next, we discuss some advanced operations that attempt to simplify and to a
certain degree, automate the discovery of performance issues.

\vspace{0.08in}
\noindent{\bf load\_imbalance}:
When parallelizing an application over a large number of processes or threads,
load imbalance across processes can lead to worse performance and limit the
highest speedup achievable. To help analyze load imbalance in a quick and easy
manner, we provide this operation, which takes as input a single metric such as
exclusive time and the number of processes to output per function that have the
highest ``load'' for that metric. The list of processes and imbalances (maximum
time across all processes / mean time per process) are provided per function,
making it easy for the user to identify which functions are especially critical
for relieving scaling bottlenecks.

\vspace{0.08in}
\noindent{\bf idle\_time}:
Processes in an MPI program often wait for messages to arrive, either in a
blocking {\tt MPI\_Recv} or {\tt MPI\_Wait}. This is referred to as ``idle
time,'' and can indicate a number of performance issues such as load imbalance,
network congestion, or system noise (OS jitter).  Reducing idle time can
improve the scaling of parallel applications. This operation calculates the
idle times per process, which can be sorted to identify the most or least idle
processes. Additionally, \pipit allows users to specify alternative operations
to qualify as ``idle time,'' to account for different programming models.

\vspace{0.08in}
\noindent{\bf pattern\_detection}:
Identifying a portion of the trace to focus on is a difficult task in trace
analysis. Typically, users manually look at the timeline to identify
``interesting'' regions to focus on, which can be cumbersome.  Automatic
pattern detection can help us find repeating patterns which can either signal
performance issues or help us find the start and end of a loop in an iterative
program. However, detecting patterns is a challenging task if attempted
manually. To simplify this task, \pipit provides the {\tt pattern\_detection}
operation. Through this operation, we also demonstrate the use of other Python
libraries such as STUMPY~\cite{law2019stumpy}, which can detect similar
repeated subsequences in time series data using matrix
profiles~\cite{matrix_profile}. This function takes a start event as an
argument that indicates where the pattern might be starting from and
automatically detects patterns.

\vspace{0.08in}
\noindent{\bf calculate\_lateness}:
For different reasons, functions in a program can be delayed beyond the
programmer's expectations. This operation computes the difference between when
a function call actually completes, and when it could theoretically have
completed. To do this, we extract the program's {\it logical structure}, which
assigns each operation a global index using a happens-before
relationship~\cite{lamport}. \pipit uses this logical structure to calculate
the {\it lateness} metric, as defined by Isaacs et al.~\cite{lateness}, to
determine how far behind the ideal execution time a function call is lagging.
We can then use this metric to identify functions in a trace that are the most
``late''.

\vspace{0.08in}
\noindent{\bf critical\_path\_analysis}:
Critical path is the longest sequence of dependent operations in a parallel
program. The duration of the operations on the critical path determines the
runtime of the execution. Therefore, it is important to optimize the operations
on the critical path to improve overall performance. To identify the critical
path, we start from the process that is the last to finish execution in a
trace. We trace back through the sequence from the last operation to the first
operation considering the messaging dependencies between processes. 

\vspace{0.08in}
\noindent{\bf multi\_run\_analysis}:
Another difficult challenge in performance analysis is the comparison of traces
from multiple executions.  For example, a user might be interested in analyzing
how their application scales with different numbers of processes.  We provide a
simple operation that takes multiple trace datasets as input and computes flat
profiles for each of them, with respect to a specific metric. If the datasets
were gathered on different numbers of processes, the result easily highlights
the performance difference of different functions across the runs. Such
scriptable analysis is impossible to do in a GUI-based setup, especially for
traces from more than two executions.

\subsection{Data Reduction}

Finally, \pipit also supports filtering the trace data by different parameters
to reduce the amount of data one wishes to analyze. A user might be interested
in analyzing the traces from a subset of processes, or for a time period
shorter than the entire execution, or for a certain set of functions.

\vspace{0.08in}
\noindent{\bf filter}:
This operation enables users to filter trace events by attributes such as name,
timestamp, and process. Users can instantiate {\tt Filter} objects and use
logical operators to create compound filters. This operation returns a new {\tt
Trace} object with a reduced DataFrame of events. All of \pipit's API
operations can be applied to this reduced trace. This operation can make a
significant difference in being able to analyze extremely large traces.

\section{Visualization Support in \pipit}
\label{sec:vis}
While \pipit is mainly designed as a Python library for programmatic trace
analysis, we also provide a basic visual interface that complements the API
operations described above. Interaction has been shown to be key in
understanding complex trace data \cite{trace-vis-task-dependencies}. Therefore,
we provide highly interactive views that can be displayed in a Jupyter notebook
using Bokeh, an open-source Python visualization library \cite{bokeh}.

\vspace{0.08in}
\noindent{\bf plot\_timeline}
displays the events in a trace over time. Function calls are shown as
horizontal bars, while instant events are shown as diamonds. MPI messages are
represented as arrows from the sender to the receiver. In order to minimize
clutter, these arrows only drawn after user interaction.  This view makes use
of several \pipit API operations described in Section~\ref{sec:api}, including
{\tt match\_caller\_callee}.

The timeline offers extensive customization depending on the type of
exploration the user is interested in. For example, users can select the time
range to display, colors for different events, and how the y-axis should be
organized (e.g. to show call stack depth, or a collapsed view of each process).
Users can add custom annotations to the timeline, such as highlighting a region
of interest or adding a path (like the critical path discussed in
Section~\ref{sec:api}).

Additionally, the timeline is designed to scale to large traces by rasterizing
tiny functions and clusters into images. When the user zooms in, the rasterized
images are replaced with the actual events on the go, so that the trace can be
explored at any granularity. This way, \pipit can generate a lag-free visual
representation of millions of events in real-time.

\begin{figure*}[t]
  \centering
    \includegraphics[width=2.3in]{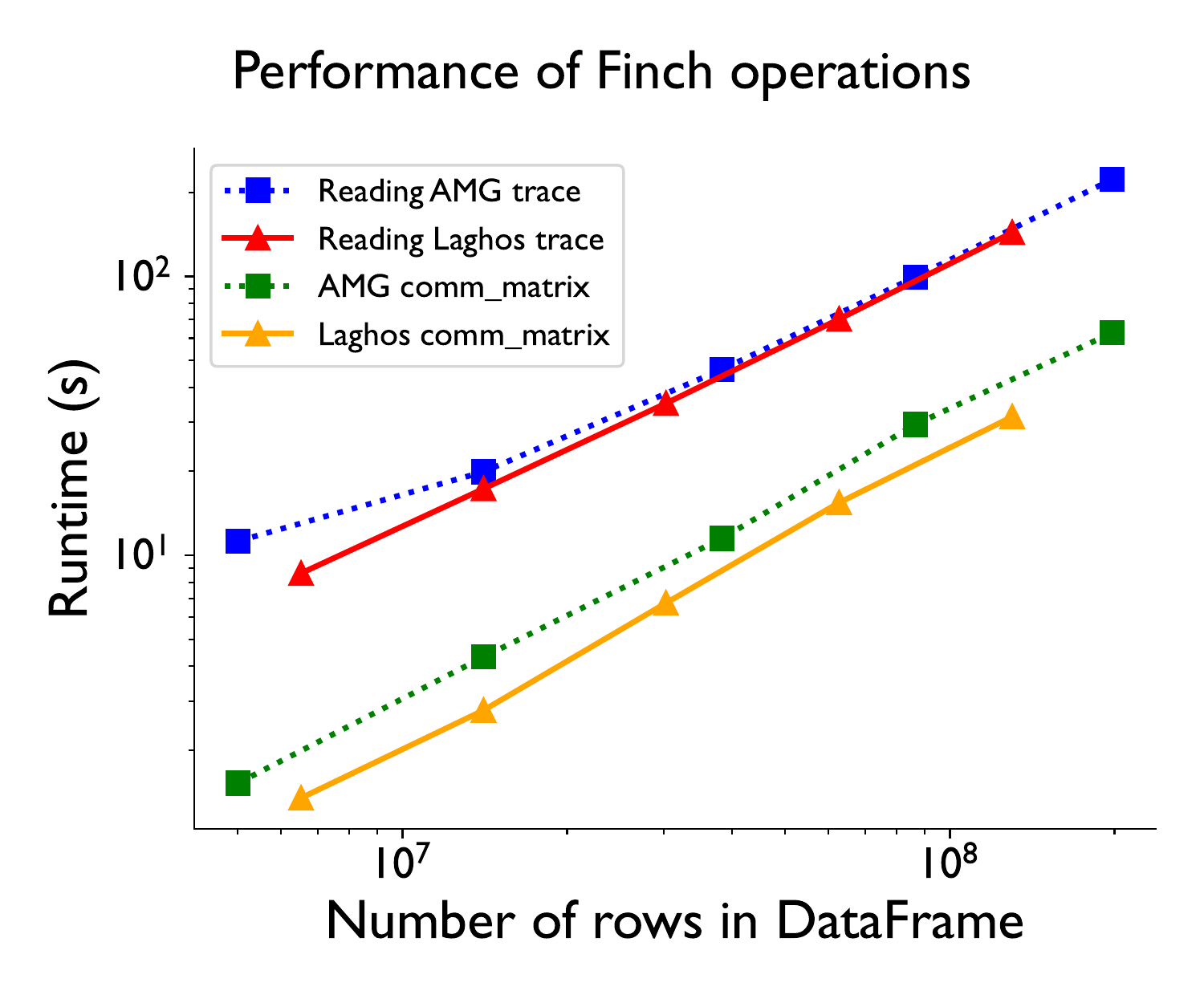}
    \includegraphics[width=2.3in]{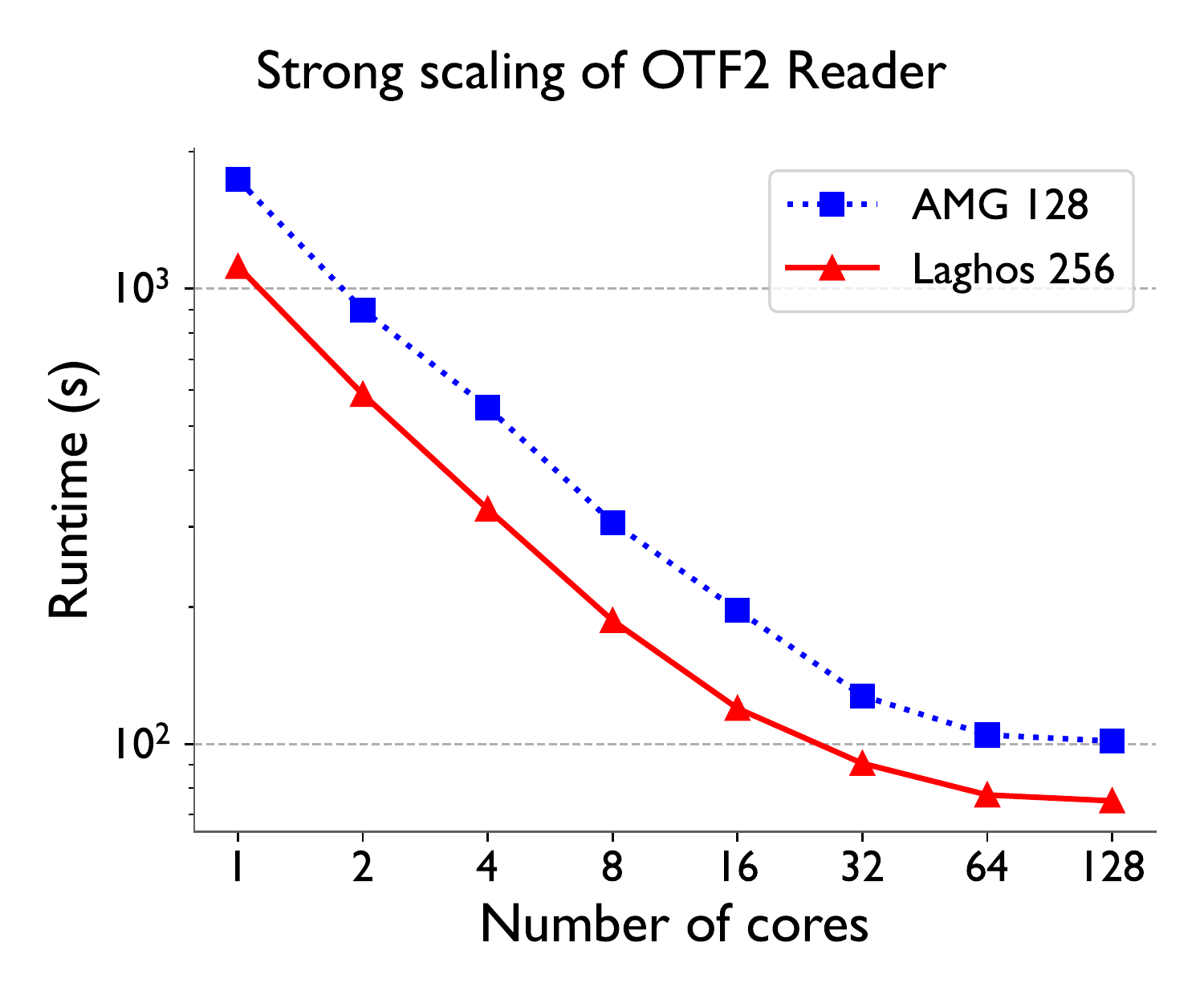}
    \includegraphics[width=2.3in]{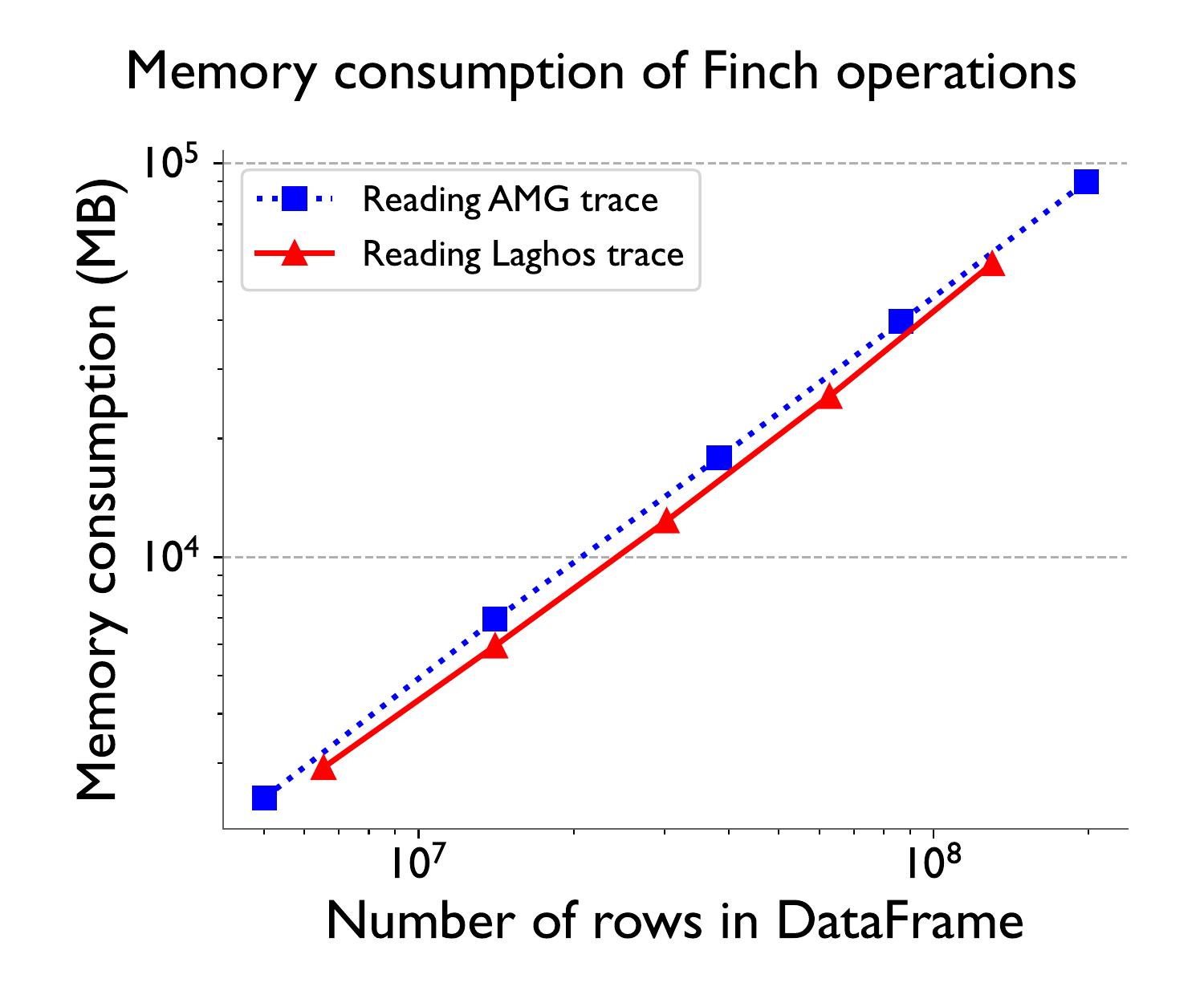}
    \caption{Performance of the OTF2 Reader and {\tt comm\_matrix} for various
traces of AMG and Laghos (left). Strong scaling performance of the OTF2 Reader
for AMG 128-process and Laghos 256-process traces (center).  Memory consumption of the OTF2
Reader for various traces of AMG and Laghos (right).}
  \label{fig:perf-figs}
\end{figure*}

\vspace{0.08in}
\noindent{\bf plot\_time\_profile}
displays the time profile of the trace, as described in Section~\ref{sec:api}
using a stacked bar graph that evolves over time.  The stacked bars are
color-coded by the function name, and their heights represent the time spent in
each function in each time bin. Figure \ref{fig:time-profile-2} shows a sample
time profile view for a Tortuga execution on 64 processes (region of interest
is in the middle).

\vspace{0.08in}
\noindent{\bf plot\_comm\_matrix}
displays the trace's communication matrix as a heatmap which encodes
communication volume using color intensity. A logarithmic scale can also be
selected to better visualize the communication patterns.

\begin{figure}[h]
    \centering
    \includegraphics[width=\columnwidth]{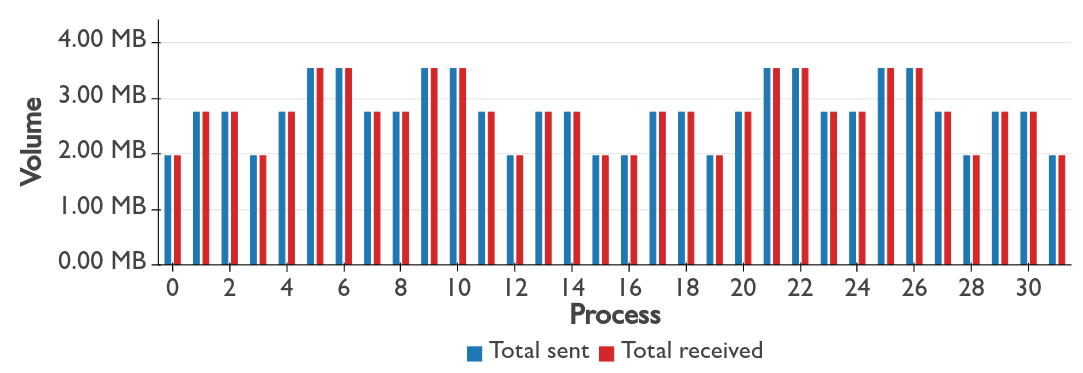}
    \caption{Communication by process view of a Kripke execution on 32 processes.}
    \label{fig:comm-summary}
\end{figure}

\vspace{0.08in}
\noindent{\bf plot\_comm\_by\_process}
displays the trace's communication volume by process as a bar graph, where the
heights of the bars represent the total message volume sent and received by
each process.  Figure~\ref{fig:comm-summary} shows this view for a Kripke
execution on 32 processes. We observe that each process can be placed in one of
three groups based on its total communication volume.

\vspace{0.08in}
We provide additional visualizations that are not described here in detail.
These views either directly correspond to an API operation in \pipit or use a
combination of multiple operations for a multi-faceted visualization.

\section{Performance of \pipit Operations}
In this section, we present the performance of a few hand-picked \pipit
operations to understand their scalability.  We first analyze the scalability
of various \pipit operations w.r.t.~increasing trace sizes. For each
experiment, we average the execution times over three trials. All the
experiments in this section were performed on a single node of an HPC cluster
with a dual 64-core AMD EPYC 7763 processor (2.45 GHz base, 3.5 GHz turbo).
Figure~\ref{fig:perf-figs} (left) shows the time spent in the OTF2 reader and
{\tt comm\_matrix} operation for AMG and Laghos traces of varying sizes. We see
that the time spent in each function scales linearly with the number of rows in
the DataFrame (size of the trace). 

Since the time for reading traces grows with trace size, we parallelized the
reading of input traces in certain file formats, including OTF2 and
Projections, using Python' multi-processing package. Figure~\ref{fig:perf-figs}
(center) shows the time spent by \pipit's OTF2 reader in reading two trace
datasets: AMG (128 processes) and Laghos (256 processes). The OTF2 reader
performance scales well with the number of cores.  We also looked at the memory
consumption of \pipit when reading traces of varying sizes.
Figure~\ref{fig:perf-figs} (right) shows the memory consumption of the OTF2
reader. We do not present the memory consumption of other API functions as they
do not increase memory usage significantly after the initial read.

\section{Case Studies}
In this section, we demonstrate the utility and power of programmatic analysis
using the \pipit API in simplifying the performance analysis of parallel
traces.  We use traces from a variety of parallel applications, including
AMG~\cite{amg-proxy-app}, Laghos~\cite{laghos-proxy-app},
Kripke~\cite{kunen2015kripke}, Tortuga (a CFD code), Loimos (a Charm++-based
epidemiology simulator), and AxoNN (a parallel deep learning
framework)~\cite{singh:ipdps2022}.

\subsection{Load imbalance analysis}

\begin{figure}[h]
  \centering
  \includegraphics[width=\columnwidth]{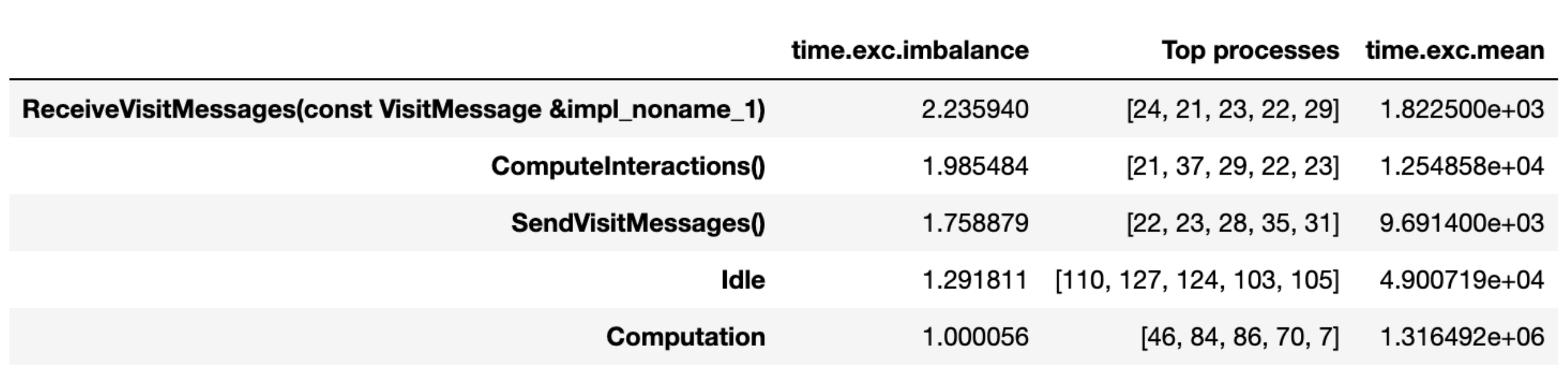}
\begin{lstlisting}
loimos_128 = finch.Trace.from_projections('loimos_128')

loimos_128.load_imbalance(num_processes=5).sort_values(by='time.exc.imbalance', ascending=False).head()
\end{lstlisting}
  \caption{Analyzing load imbalance in the five most time consuming functions in a Loimos 128-process trace.}
  \label{fig:load-imbalance}
\end{figure}

Load imbalance is a commonly occurring performance problem, and we can use the
\pipit load\_imbalance function to expose asymmetry in aggregated runtimes of
functions across processes. The code in Figure~\ref{fig:load-imbalance}
demonstrates such an example, where a Projections trace of Loimos, an epidemic
simulation framework, is used. As we can see, with a few lines of Python code,
the output of the {\tt load\_imbalance} operation can be filtered to identify
the load imbalance in the five most time consuming functions.

We notice some interesting observations in the data returned by the
load\_imbalance operation. First, {\tt computeInteractions()}, which determines
which individuals get infected, is the most time consuming function.  It also
appears to have high load imbalance, second only to {\tt
ReceiveVisitMessages()}, which is a message processing function.  Another
interesting observation is that the most overloaded processes are common across
the top three functions (21, 22, 23, 29).

\subsection{Filtering to focus on subset of trace data}

As mentioned before, traces can be large when programs are traced for a long
time or for many processes or threads. In most cases, programmers want to
sub-select a portion of the trace that is ``interesting'' and might reveal
performance issues. Here, we demonstrate how two different \pipit operations
can be used to filter a trace to a more manageable size.

\begin{figure}[h]
    \centering
    \includegraphics[width=\columnwidth]{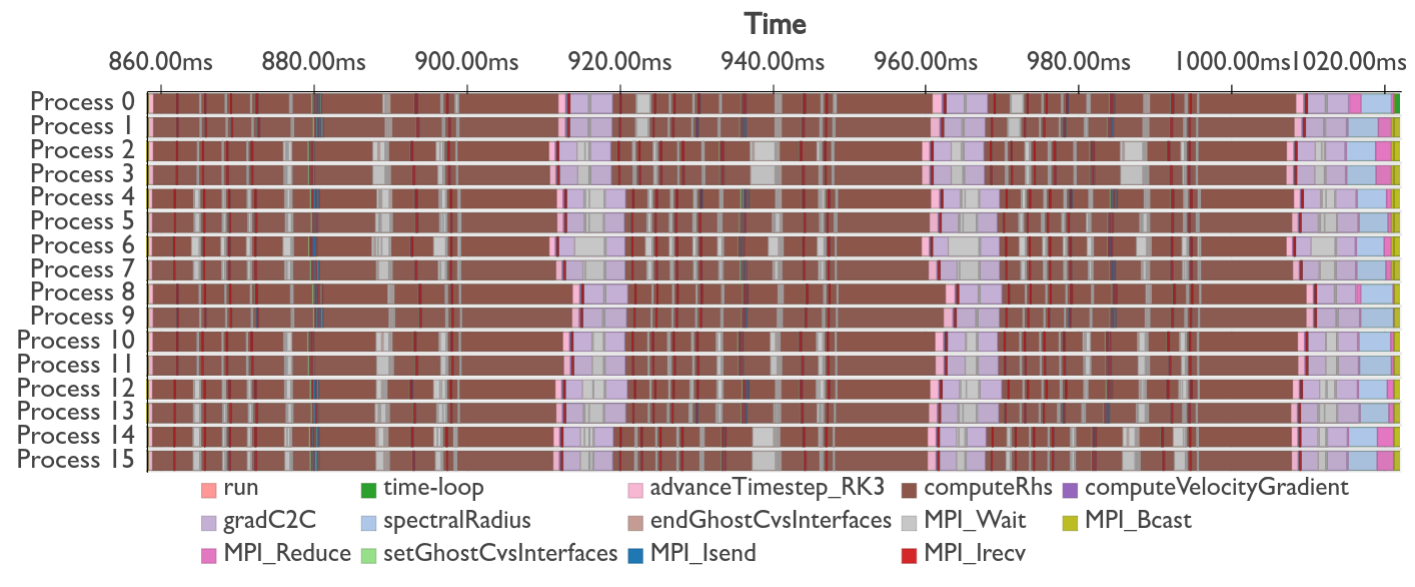}
\begin{lstlisting}
tor_16 = finch.Trace.from_otf2('./tortuga_16')
patterns = tor_16.detect_pattern(start_event='time-loop')

start = patterns[0].iloc[0]['Timestamp (ns)']
end = patterns[0].iloc[-1]['Timestamp (ns)']
tor_16.plot_timeline(x_start=start, x_end=end)
\end{lstlisting}
    \caption{Timeline view of one iteration of a Tortuga execution after detecting patterns and filtering by time.}
    \label{fig:pattern_timeline}
\end{figure}

\vspace{0.07in}
\noindent{\bf Pattern detection to identify iterations}:
We can use the results of \pipit functions such as {\tt pattern\_detection} to
filter the trace DataFrame by time. To demonstrate this, we use an OTF2 trace
of the Tortuga application executed on 16 processes. In the code listing in
Figure~\ref{fig:pattern_timeline}, we use the output of the {\tt
pattern\_detection} operation to filter the timeline by a time range, which
allows us to focus on one iteration of Tortuga.

The ability to detect patterns and identify start and end time ranges of loop
iterations can be extremely useful. When traces get large, and visualizing them
in a timeline becomes challenging, we can use the start and end of loop
iterations to filter the trace and visualize a smaller time range.

\vspace{0.07in}
\noindent{\bf Idle time to identify process outliers}:
As a complement to filtering traces by time ranges, we can also filter traces
by process IDs or ranks. We can achieve this via the {\tt idle\_time} function
in \pipit. Returning the \textit{k} most and least idle processes, {\tt
idle\_time} allows users to effortlessly identify which processes are the most
under- and over-utilized.  We use a 64-process Loimos trace to highlight the
utility of the {\tt idle\_time} operation in \pipit.

\begin{figure}[h]
  \centering
  \includegraphics[height=2.4cm]{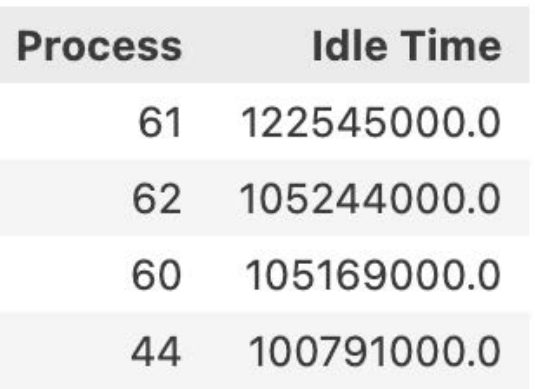}
  \hspace{0.05cm}
  \includegraphics[height=2.4cm]{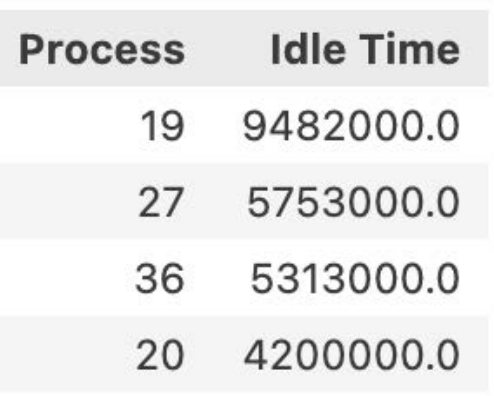}
  \includegraphics[width=\columnwidth]{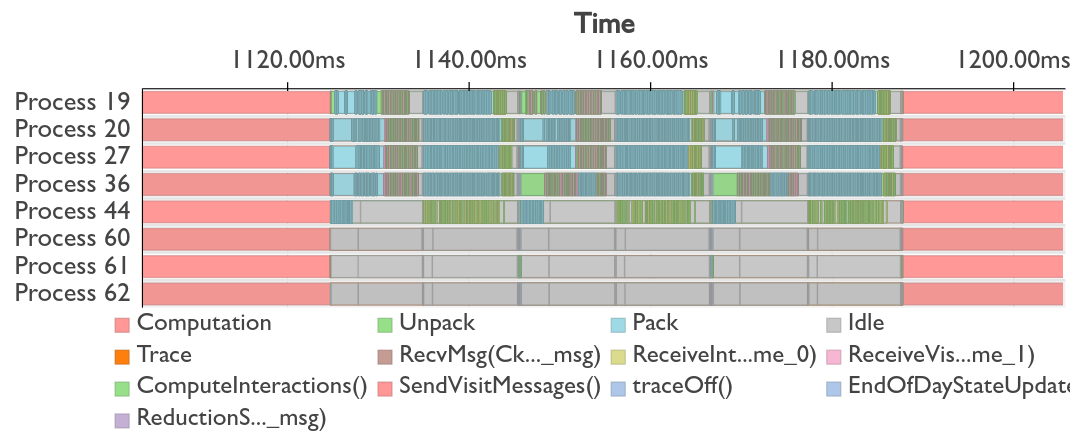}
\begin{lstlisting}
loimos_64 = finch.Trace.from_projections('./loimos_64')
idle_times = loimos_64.idle_time().sort_values(by='Idle Time', ascending=False)

most_idle_procs = idle_times['Process'].head(4)
least_idle_procs = idle_times['Process'].tail(4)

loimos_64.filter('Process', 'in', least_idle_procs + most_idle_procs).plot_timeline()
\end{lstlisting}
  \caption{Most idle processes (top-left), least idle processes (top-right), and timeline view, filtered by the most and least idle processes (bottom), of a 64-process Loimos trace.}
  \label{fig:loimos-idle}
\end{figure}

Figure~\ref{fig:loimos-idle} shows the code for calculating idle time and for
reducing the data to 8 processes in the Loimos trace, as well as the resulting
DataFrames and timeline plot. The top left and right DataFrames in the figure
show the most and least idle processes respectively.  The output of idle time
is then used to filter the trace by specific MPI ranks, reducing the size of
the DataFrame and the scope of analysis. While not necessary when using \pipit
programmatically, we opt to display the reduced dataset via the timeline to
convey the data reduction through a visually friendly medium.
Figure~\ref{fig:loimos-idle} (middle) displays the resulting timeline plot.
Evidently, this helps us easily compare the outlier processes, and clearly see
the differences in activity between processes that are the most idle and those
that are the least idle.

\subsection{Identifying communication issues}

Often times, communication is one of the primary bottlenecks in the scaling
performance of parallel codes. The next two case studies demonstrate the use of
the \pipit API to identify communication-related performance issues.

\begin{figure}[h]
  \centering
  \includegraphics[width=0.72\columnwidth]{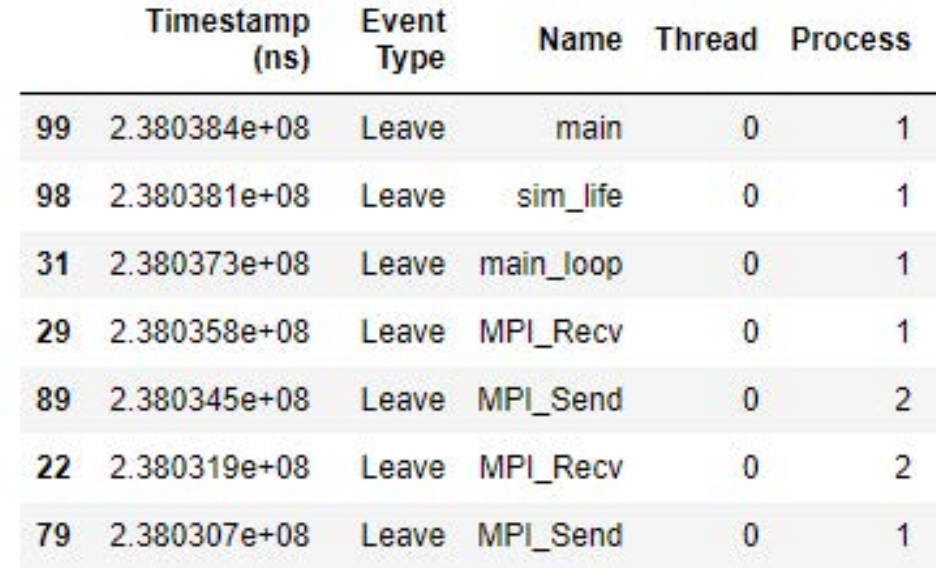}
  \includegraphics[width=\columnwidth]{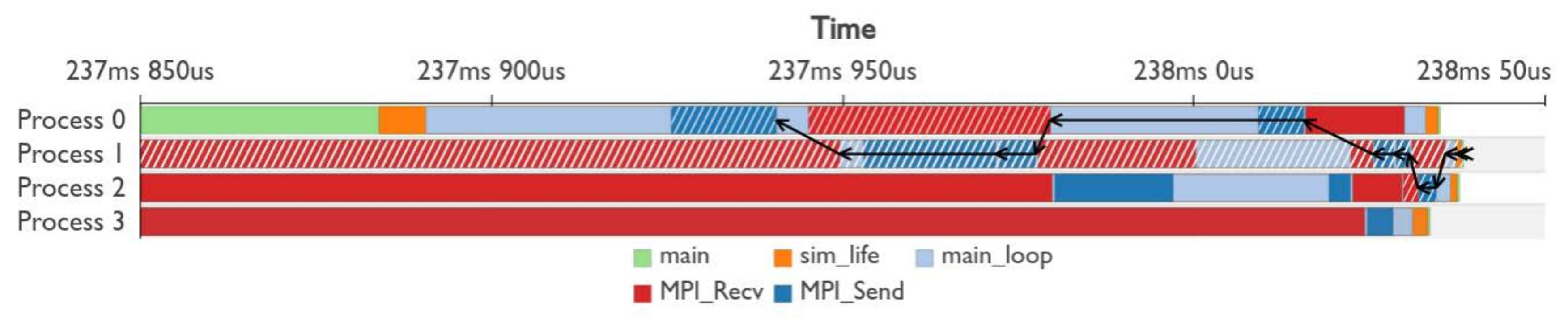}
\begin{lstlisting}
gol_4 = finch.Trace.from_otf2('./gol_4')
critical_paths = gol_4.critical_path_analysis()
display(critical_paths[0].head(7)) # dataframe

gol_4.plot_timeline(show_critical_path=True)
\end{lstlisting}
  \caption{Critical path in a 4-process trace of an MPI Game of Life program (top), and its visualization using \pipit's timeline view (bottom).}
  \label{fig:critical-path}
\end{figure}

\vspace{0.07in}
\noindent{\bf Detecting critical paths}:
We identify the critical path in a 4-process trace of a parallel Game of Life
program using the {\tt critical\_path\_analysis} function.  As shown in
Figure~\ref{fig:critical-path}, \pipit outputs a critical path as a dataframe,
which can then be provided to the plot\_timeline function to visualize critical
paths. The code block in Figure~\ref{fig:critical-path} demonstrates how to
obtain a critical path dataframe and visualize it. As can be seen, the critical
path starts from the last process to exit the main function and traces
backwards in time until it finds a dependent operation, which is {\tt
MPI\_Recv} in this case. From there, it identifies the corresponding {\tt
MPI\_Send} on the sender process and continues doing this until it finds
another dependent operation. The critical path ends on Process 0, which appears
to spend more time on the main function.  Process 1 waits for Process 0 to
complete its {\tt MPI\_Send}.  Optimizing the tasks that Process 0 performs
before its first {\tt MPI\_Send} will likely improve the overall performance.

\vspace{0.07in}
\noindent{\bf Identifying lateness using logical time}:
We can also use \pipit to identify lateness in a program's execution.
Lateness, as described in Section~\ref{sec:api}, quantifies how much each
function call lags behind its expected time of completion. We analyze lateness
in an 8-process trace of a parallel Game of Life program.

\begin{figure}[h]
  \centering
  \raisebox{-0.5\height}{\includegraphics[width=0.72\columnwidth]{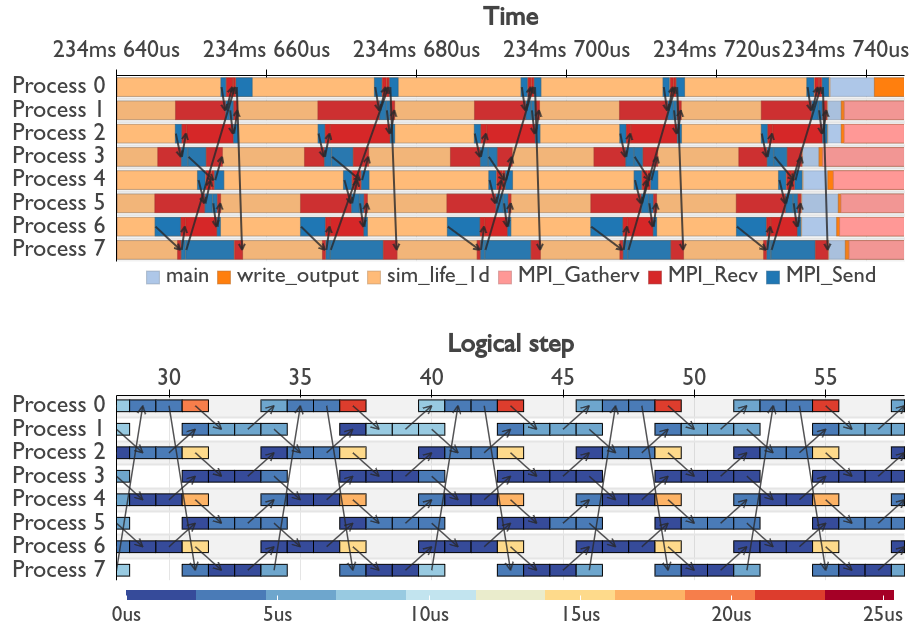}}
  \raisebox{-0.5\height}{\includegraphics[width=0.26\columnwidth]{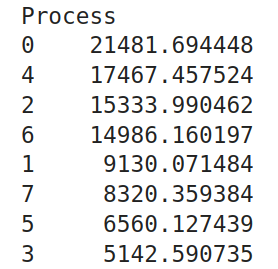}}
\begin{lstlisting}
gol_8 = finch.Trace.from_otf2('./gol_8')
gol_8.calculate_lateness()

gol_8.plot_timeline()
gol_8.plot_logical_timeline(color_by='lateness')

gol_8.groupby('Process')['Lateness'].max().sort_values(ascending=False)
\end{lstlisting}
  \caption{Regular timeline of an 8-process MPI Game of Life program with message arrows (top-left);
  its logical timeline, with each operation colored by lateness (center-left); and the maximum lateness
  for each process (right).}
  \label{fig:gol-timeline}
\end{figure}

First, we plot the regular timeline of the trace to get an overview of the
function calls (Figure~\ref{fig:gol-timeline}, top). After calculating the
lateness for each function call, we can plot the logical timeline of the trace
colored by lateness (Figure~\ref{fig:gol-timeline}, center-left).  This starts
to reveal some patterns. For instance, the {\tt MPI\_Send} calls of processes 0
and 4 consistently lag behind in each iteration, perhaps indicating load
imbalance or network issues. However, as stated by Isaacs et
al.~\cite{lateness}, such visualizations may not be enough to identify the root
cause of the lateness. To understand exactly which processes are responsible
for the lateness, we can aggregate the lateness metric of each process, as seen
in the right image of Figure~\ref{fig:gol-timeline}. This list of processes
shows us exactly which processes are late, and by how much.

\subsection{Comparing multiple executions programmatically}
\label{sec:multirun}

One of the most complex tasks in performance analysis of parallel programs is
comparing the performance of multiple executions (from scaling studies, or
changing the input problem etc.) Below, we show two examples of the use of
\pipit functions in significantly simplifying such tasks by scripting them
using Python.

\begin{figure*}[t]
  \centering
  \parbox{0.49\textwidth}{
    \includegraphics[width=\columnwidth]{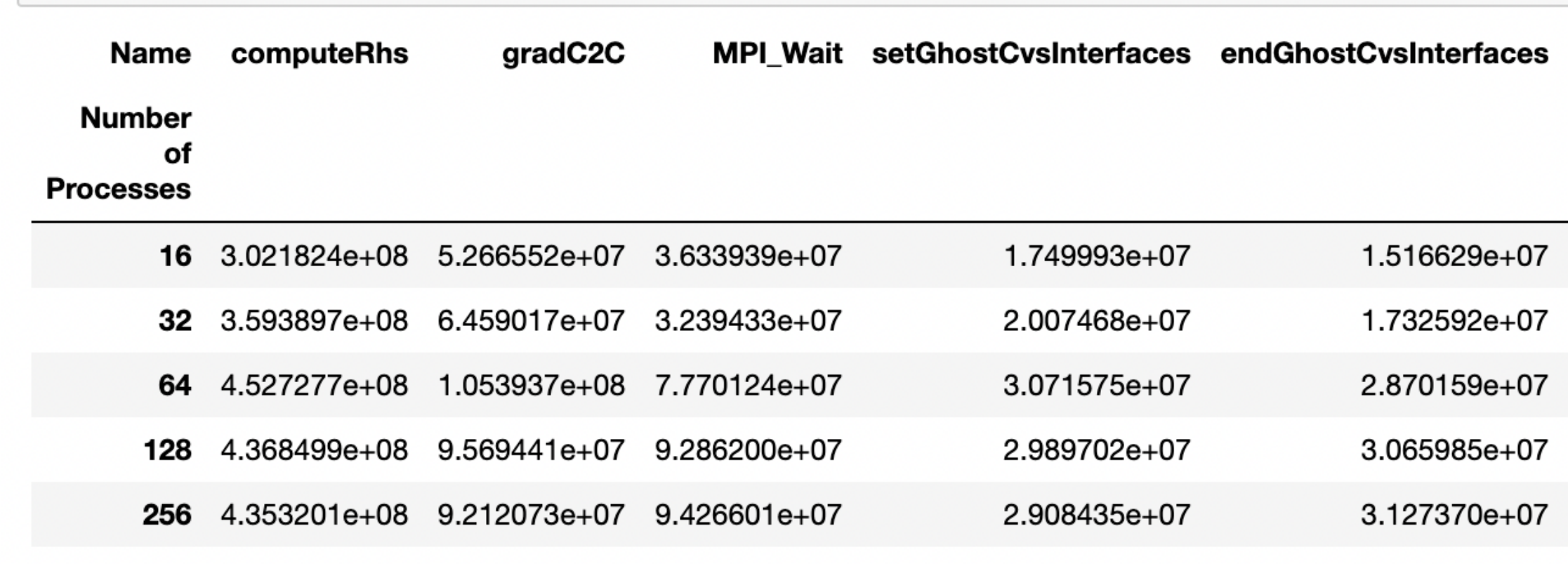}
\begin{lstlisting}
traces = [finch.Trace.from_otf2('./tortuga/'+ str(size)) for size in os.listdir('.')]
multirun_df = finch.Trace.multirun_analysis(traces)

multirun_df.plot.bar(stacked=True, ax=ax, legend=True)
\end{lstlisting}
  }
  \parbox{0.49\textwidth}{
    \centering
    \includegraphics[width=2.65in]{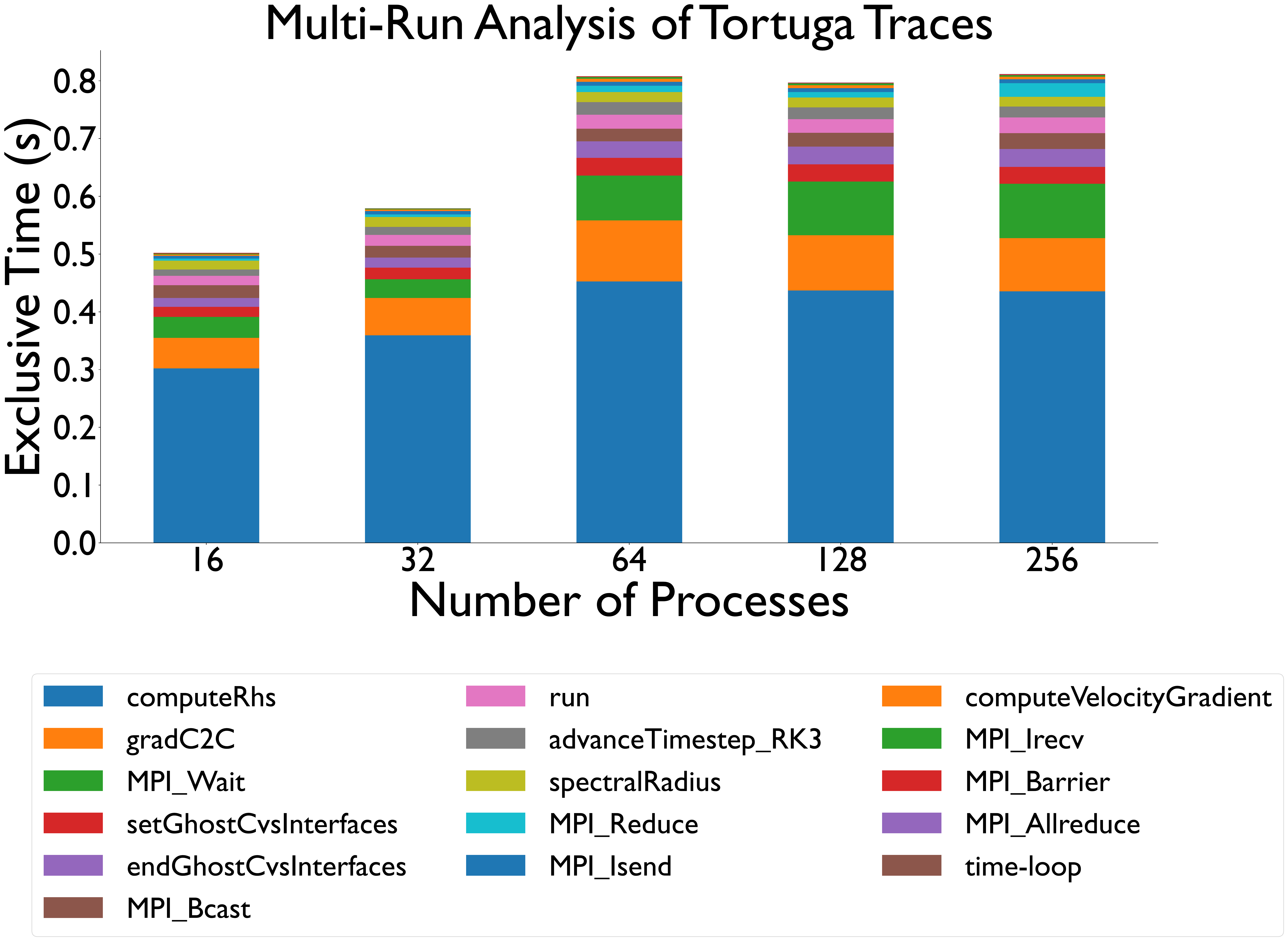}
  }
  \caption{Outcome of \pipit's {\tt multirun\_analysis} operation, executed on four instances of Tortuga (left), and its Matplotlib visualization using pandas' built-in plotting function (right).}
  \label{fig:multirun-tortuga}
\end{figure*}

\begin{figure}[h]
  \centering
  \includegraphics[width=0.85\columnwidth]{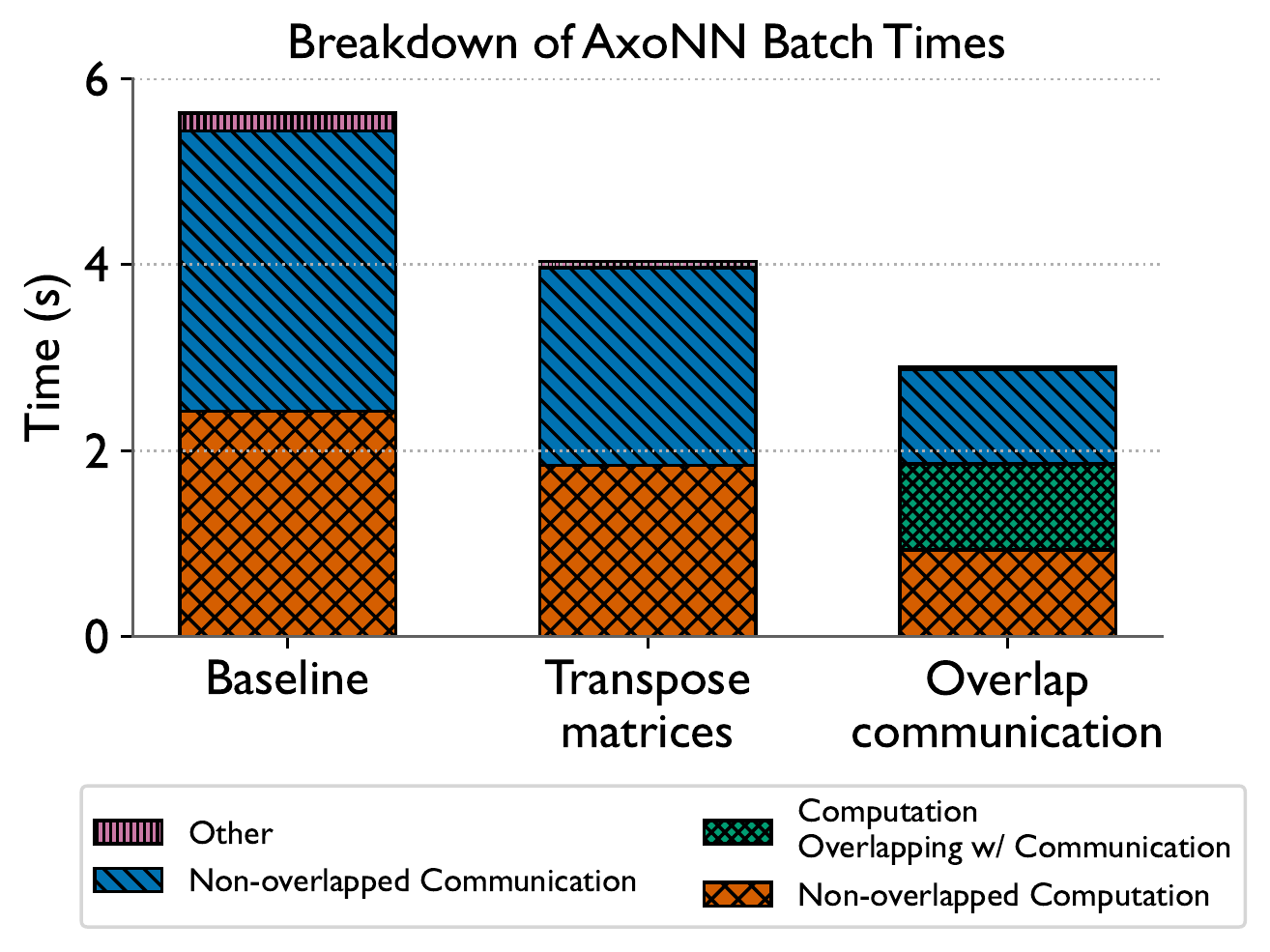}
\begin{lstlisting}
traces = [finch.Trace.from_pytorch('./'+trace_name) for trace_name in os.listdir('.')]
trace_breakdowns = [trace.comm_comp_breakdown() for trace in traces]

breakdown_list = [[phase_time for phase_time in breakdown['Average'].values()] for breakdown in trace_breakdowns]
breakdown_df = pd.DataFrame.from_records(breakdown_list)

breakdown_df.plot.bar(stacked=True, ax=ax, legend=True)
\end{lstlisting}
  \caption{Use of \pipit to analyze multiple executions of AxoNN to compare the differences in communication time and computation-communication overlap.}
  \label{fig:comm_comp_overlap}
\end{figure}

\vspace{0.07in}
\noindent{\bf Assessing communication-computation overlap}:
Analyzing the communication-computation breakdown in asynchronous applications
is crucial for optimizing system utilization. \pipit's {\tt
comm\_comp\_breakdown} operation helps identify how much time is spent in
overlapped communication as well as non-overlapped communication and
computation. We demonstrate the utility of this function by analyzing traces of
a parallel deep learning framework, AxoNN~\cite{singh:ipdps2022}. AxoNN uses
PyTorch and offloads most of its computation to GPUs. We collect PyTorch traces
for three different versions of AxoNN, optimized to different degrees.  The
results are shown in Figure~\ref{fig:comm_comp_overlap}, where we can compare
the breakdown of the per iteration times for the three executions.

In the second execution, unnecessary communication is avoided by changing data
layouts (transposing matrices) and in the third execution, various
communication routines are overlapped with computation to improve performance.
The performance improvements in the second and third execution are clearly
evidenced by Figure~\ref{fig:comm_comp_overlap}, where we can see the
significant reduction in communication time and the increase in overlap in the
third execution. To the best of our knowledge, creating such summary plots from
multiple GPU-based executions is not possible with any other performance tool
currently. Moreover, as is evident in the code listing, \pipit makes this
relatively easy, only requiring a few lines of Python code.

\vspace{0.07in}
\noindent{\bf Identifying scaling issues}:
We can use the \pipit function {\tt multi\_run\_analysis} to identify which
functions scale poorly as we run on more processes. Such analysis is
painstakingly difficult to do with traditional GUI-based tools. With a few
lines of Python code, we can compute flat profiles over several datasets,
resulting in a DataFrame as shown in Figure~\ref{fig:multirun-tortuga} (left).
This analysis is done using traces collected from five Tortuga executions on 16
to 256 processes.

We plot the output of the {\tt multi\_run\_analysis} operation using
Matplotlib, as shown in Figure~\ref{fig:multirun-tortuga} (right). The
execution time increases significantly when scaling Tortuga from 32 to 64
processes. {\tt computeRhs} and {\tt gradC2C} account for a significant portion
of the program's time and increase the most, so they are likely the scalability
bottlenecks. Both of these are computationally heavy as they compute
time-derivatives ({\tt computeRhs}) and gradients ({\tt gradC2C}) on a
three-dimensional tensor. As we can see, using this approach, a user can easily
compare different traces with \pipit and identify functions to focus on when
optimizing their application. Incidentally, this also shows how users can
easily leverage other Python libraries in conjunction with \pipit.

\section{Conclusion}
In this paper, we present a new Python-based performance analysis tool called
\pipit for analyzing parallel execution traces. Through \pipit's design and
implementation, we sought to solve the following challenges:~(1)~Support
several file formats in which traces are collected, to provide users with a
unified interface that works with outputs of many different tracing
tools.~(2)~Provide a programmatic API, which allows users to write simple code
for trace analysis and provides several benefits such as flexibility of
exploration, extensibility, scalability, and reproducibility of workflows.
And~(3)~Automate common performance analysis tasks for analyzing single and
multiple executions. To the best of our knowledge, \pipit is unique in its
capabilities in terms of supporting several file formats, providing a
programmatic API, and several operations to identify performance issues easily,
and to some extent, automatically.  We believe that \pipit can revolutionize
how HPC developers and performance engineers analyze the performance of their
codes, and that it will improve the efficiency of both parallel programs and
HPC programmers. We hope that other analysis tools will be developed on top of
\pipit.


\bibliographystyle{IEEEtran}
\bibliography{bib/cite,bib/pssg}


\end{document}